\documentclass[a4paper,12pt]{article}
\usepackage{latexsym,amsmath,amssymb,amsbsy,graphicx, subcaption}
\usepackage[justification=centering]{caption}
\usepackage[utf8]{inputenc}
\usepackage[english]{babel}
\usepackage{float}
\usepackage[left=15mm,right=15mm,top=2cm,bottom=2cm,bindingoffset=0cm]{geometry}
\DeclareUnicodeCharacter{2212}{-}
\graphicspath{ {./images/} }

\title{From Exponential to Gaussian Tails: Fractal Wavefront Scaling and the $\kappa$-Weibull Distribution at Phase Transitions}
\date{July 2025}
\begin{document}
\newpage
\begin{center}
{\Large{From Exponential to Gaussian Tails: Fractal Wavefront Scaling and the $\kappa$-Weibull Distribution at Phase Transitions}}\\[9pt]		

{\large O.V. Pavlovsky$^{1,2 a}$, M.Yu. Satleikin$^{1,2 b}$}\\[6pt]

\parbox{.96\textwidth}{\centering\small\it
        $^1$ Kurchatov Institute, 1 Kurchatov Square, Moscow, 123182 Russia\\
	$^2$ Lomonosov Moscow State University, Leninskiye Gory 1-3, Moscow, 119991 Russia\\
	E-mail: $^a$pavlovsky@physics.msu.ru,
$^b$mihailsatleykin03@gmail.com}\\[1cc]
\end{center}

{\parindent5mm

{\footnotesize The paper examines the features of critical evolution in an active medium modeled by a cellular automaton. The system evolves according to stochastic rules, exhibiting two qualitatively distinct dynamical regimes: one of fading activity and another of all-filling activity waves. Each regime is characterized by its own fractal dimension and fluctuation statistics, both of which are analyzed in detail. Within the critical interval, the system displays a nontrivial fractal dimension and a fluctuation distribution that bridges the two regimes. This bridging behavior arises because, in the critical interval, the dynamics are governed by avalanches.}
\vspace{2pt}\par}

\small Keywords: critical phenomena, self-organization, cellular automata

\section{Introduction}
This article investigates wave propagation in an active medium \cite{Engelbrecht}, \cite{Klimontovich}, \cite{Nekorkin}, which modeled using a stochastic cellular automaton \cite{Pavlovsky}. The automaton is defined on a planar regular square lattice, where each cell can be in one of three states: sensitive, active or immune. This terminology can be motivated by analogy with epidemiological models \cite{Kærn}, \cite{Gizzi}, \cite{Sidorova}, \cite{Tverdislov}. The system evolves according to the following rules:
\begin{itemize}
\item An active cell becomes immune in the next time step.
\item An immune cell remains immune for $M$ time steps and then transitions to the sensitive state.
\item A sensitive cell may be independently excited by each active neighbor with probability $G$.
\end{itemize}
\begin{figure}[H]
    \centering
    \subfloat[]{\includegraphics[scale = 0.1915]{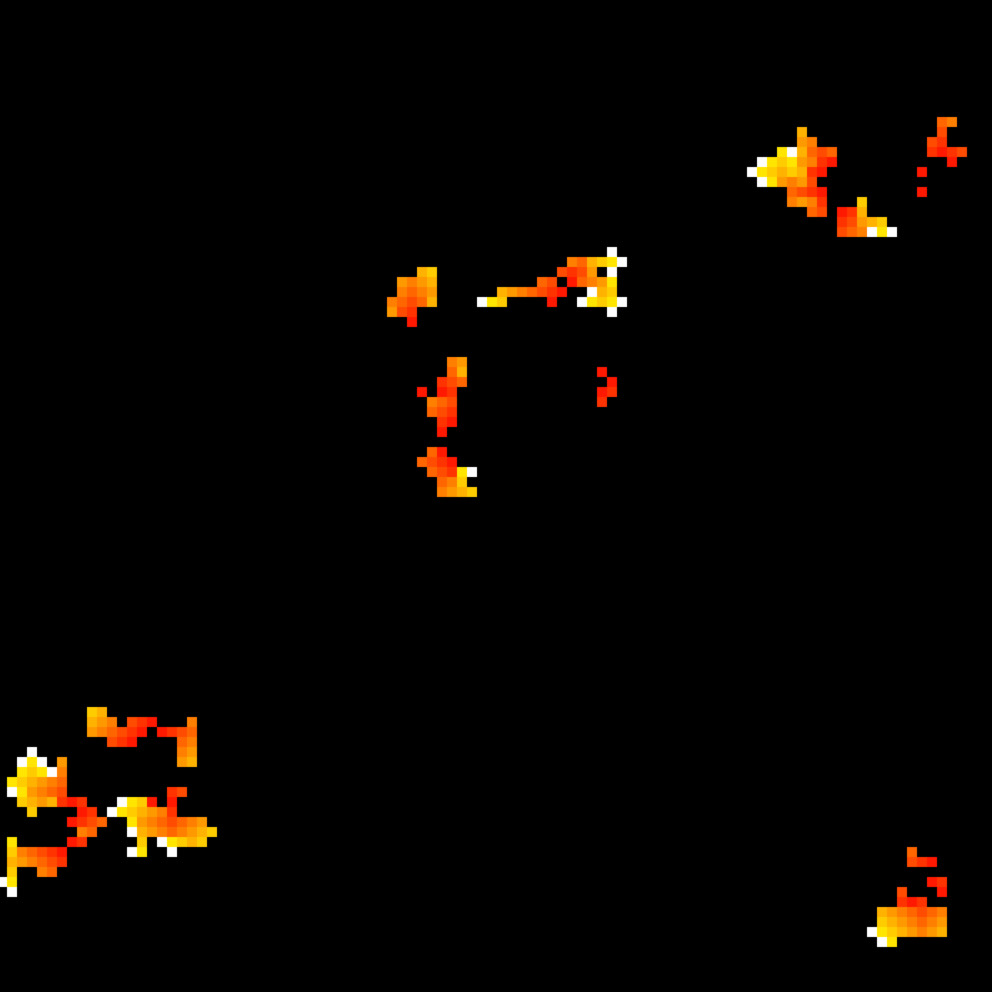}}
    \centering
    \subfloat[]{\includegraphics[scale = 0.19]{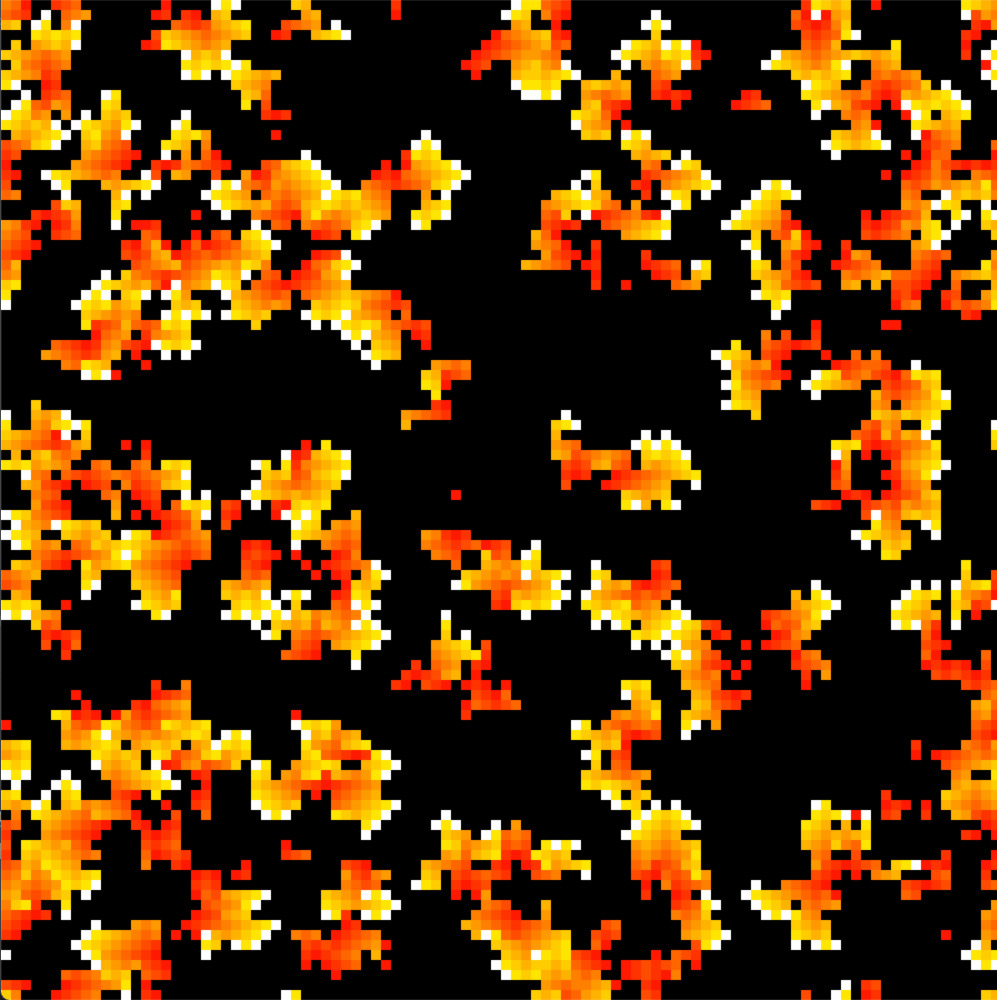}}
    \centering
    \subfloat[]{\includegraphics[scale = 0.19]{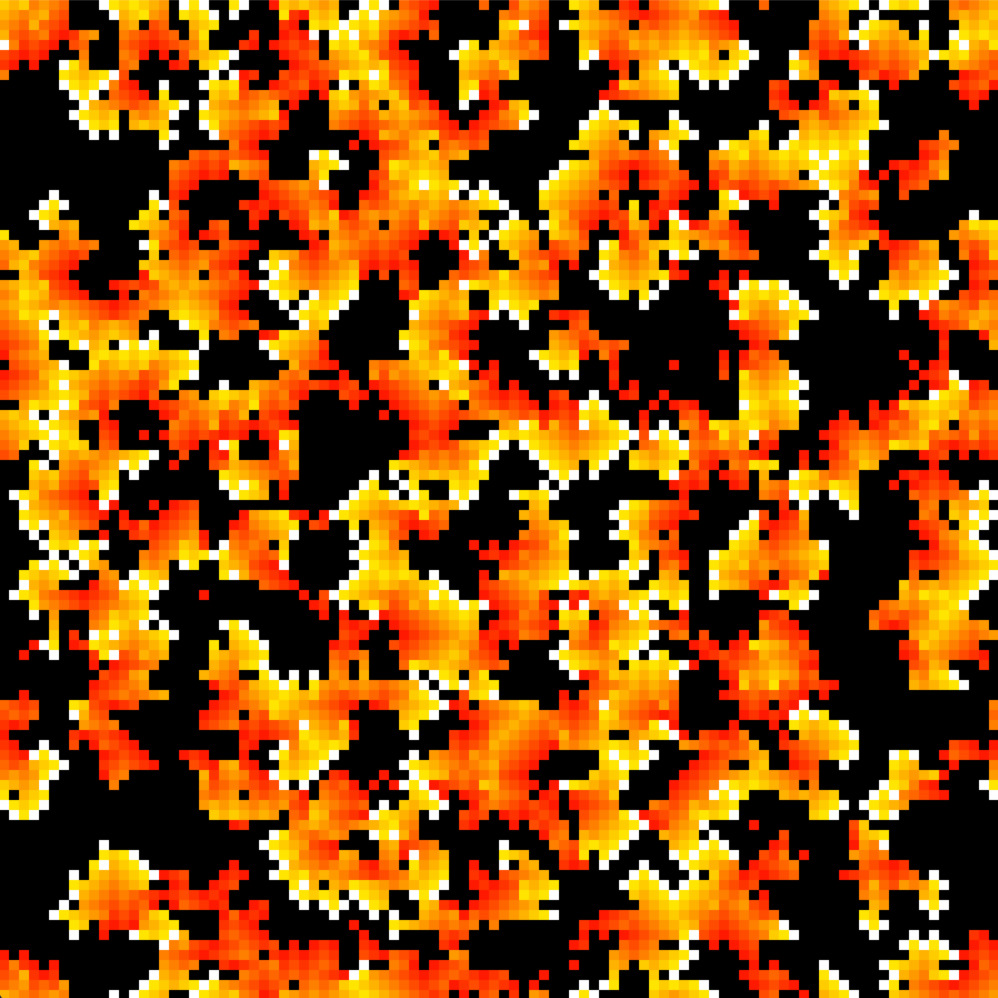}}
    \centering
    \subfloat[]{\includegraphics[scale = 0.19]{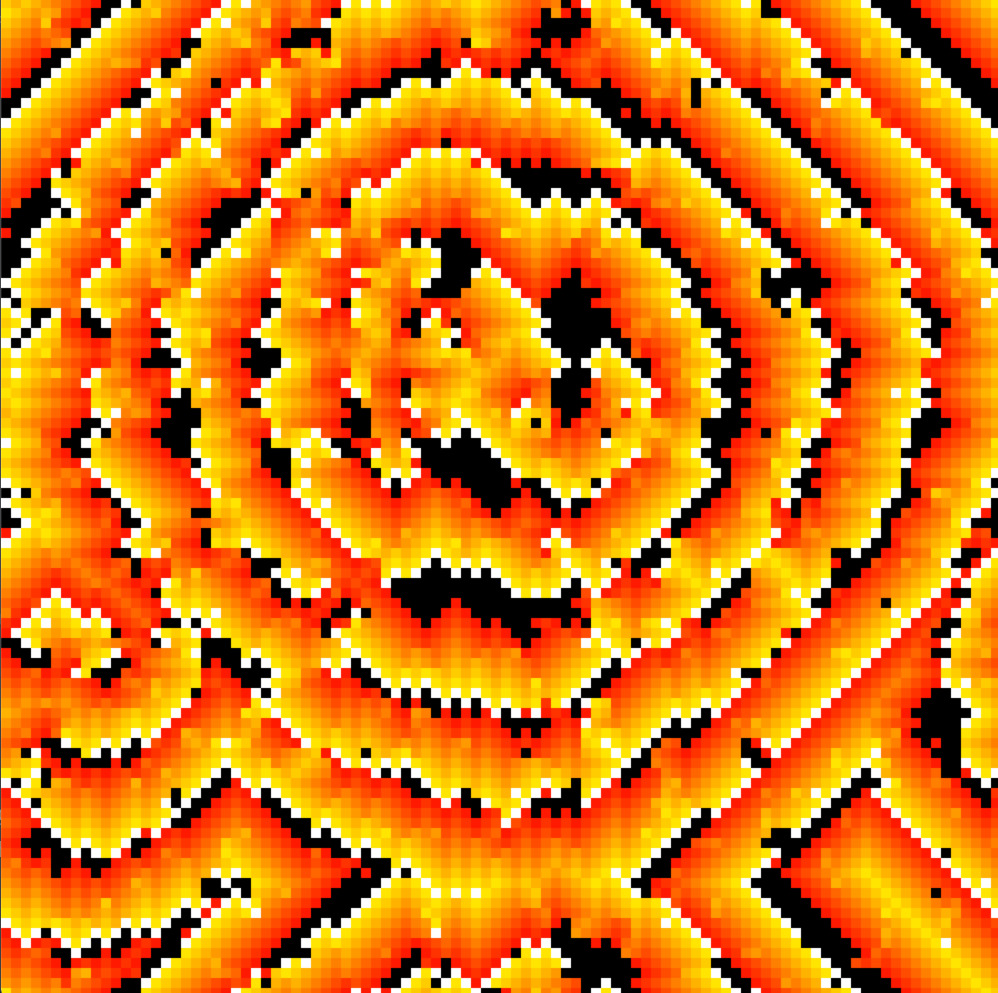}}\\
    \centering
    \subfloat[]{\includegraphics[scale = 0.19]{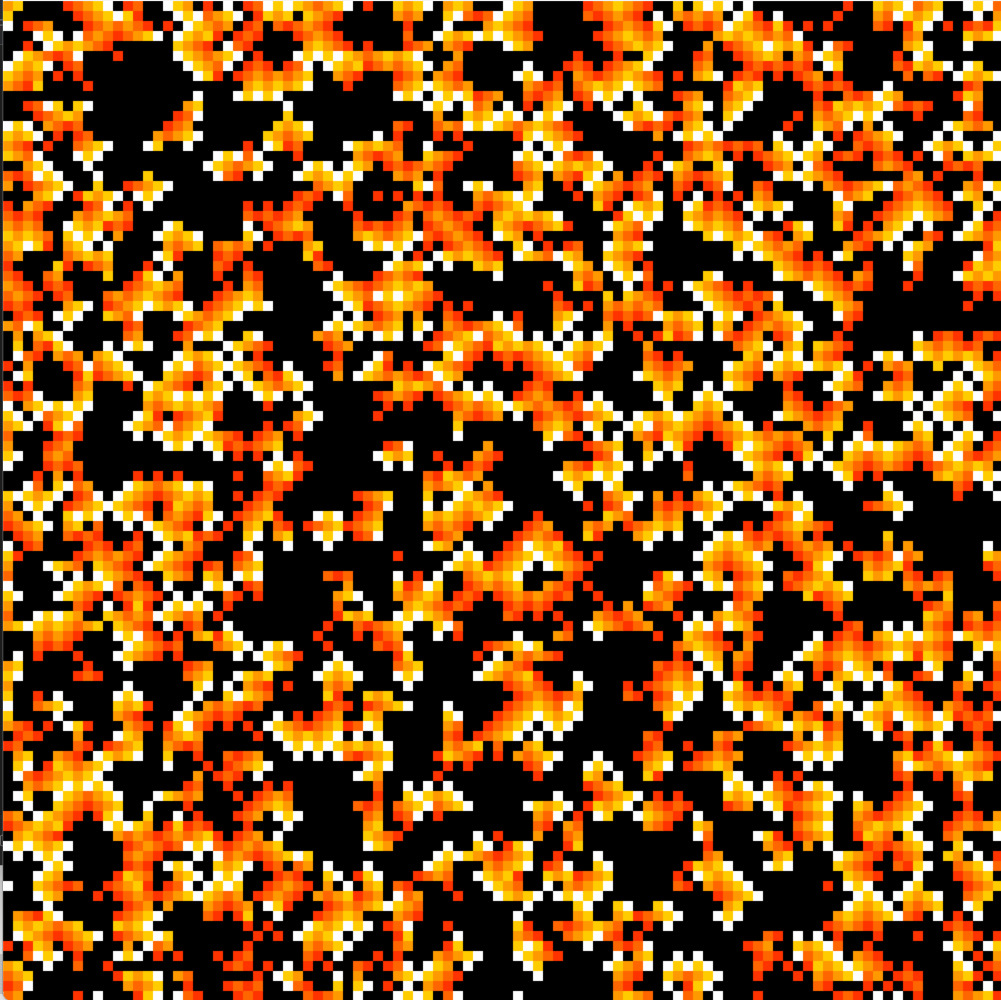}}
    \centering
    \subfloat[]{\includegraphics[scale = 0.19]{G06_M10.jpg}}
    \centering
    \subfloat[]{\includegraphics[scale = 0.19]{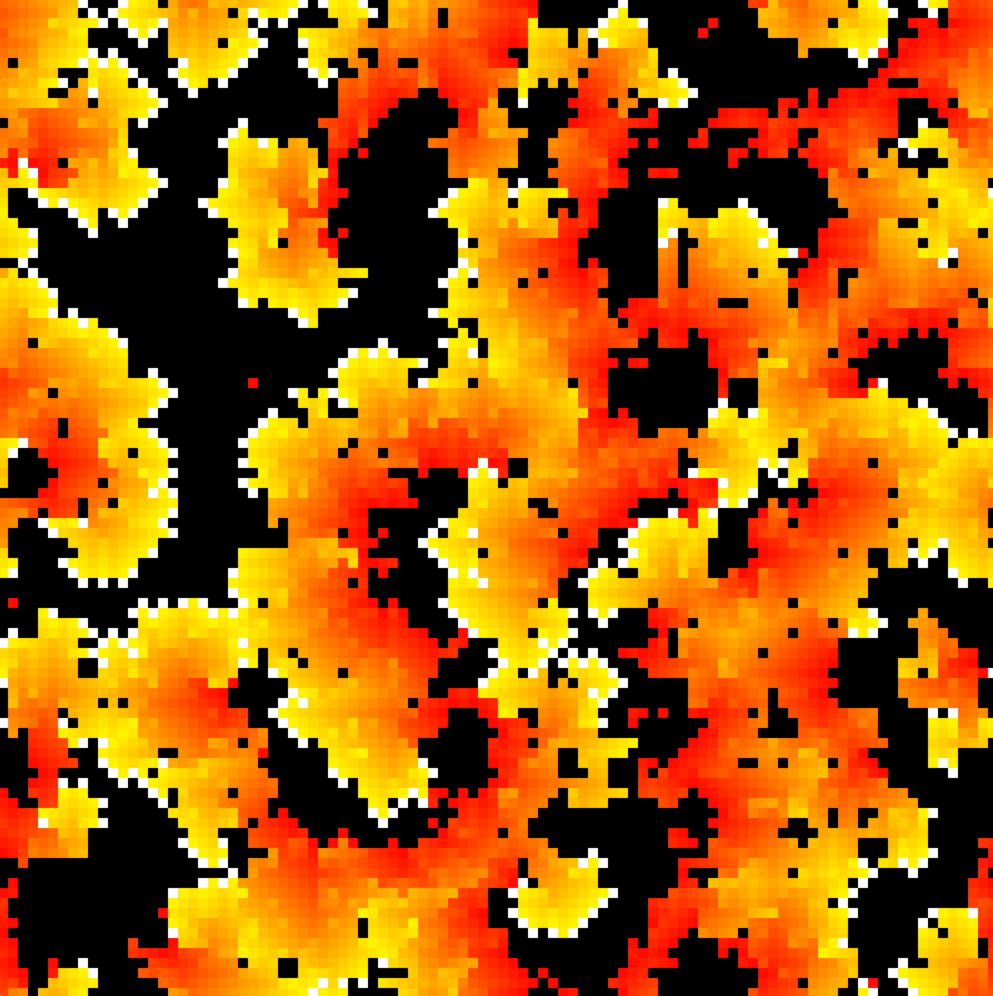}}
    \centering
    \subfloat[]{\includegraphics[scale = 0.19]{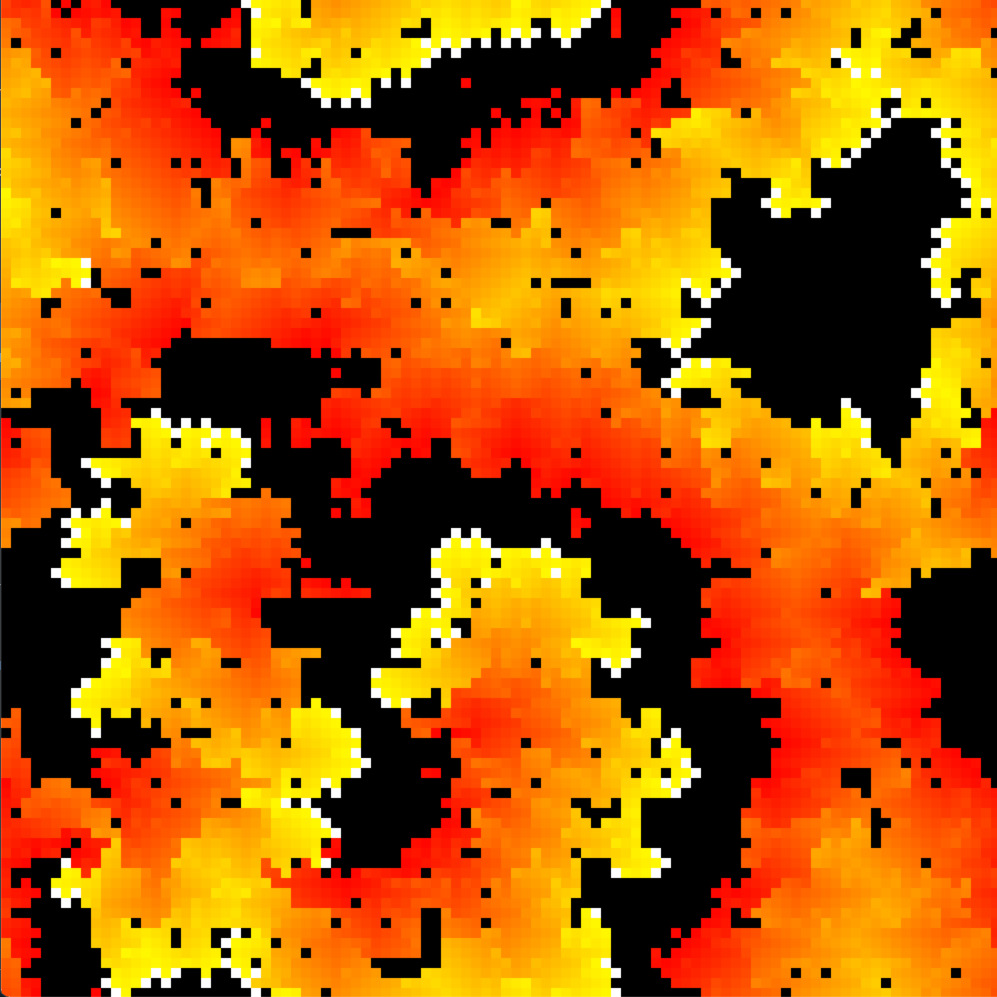}}
    \caption{The dynamic equilibrium, \ref{Dyn}a – $G = 0.45$, \ref{Dyn}b – $G = 0.5$, \ref{Dyn}c – $G = 0.6$, \ref{Dyn}d - $G = 0.8$, \ref{Dyn}e – $M = 5$, \ref{Dyn}f – $M = 10$, \ref{Dyn}g – $M = 20$, \ref{Dyn}h - $M = 35$}
    \label{Dyn}
\end{figure}
The automaton exhibits two qualitatively modes of evolution. When the activation probability lies below a critical threshold, all activity in the system fades out. Above this threshold, it organizes into waves that eventually fill the entire space, driving the system toward thermodynamic equilibrium (Fig. \ref{Dyn}). In this equilibrium mode each active cell produces on average the same number of active cells in the subsequent time step. This property gives rise to an invariant of equilibrium that is independent of the lattice geometry. The invariant is the average number of active cells generated by a single active cell which has mean number of sensitive neighbors. Dynamic equilibrium is attained when this invariant equals one. Based on this we can determine the conditions under which it is possible and derive a simple estimate of the critical probability.\\

The equilibrium activity wavefront forms a stochastic fractal and its structure scales with $M$ (Fig. \ref{Dyn}e -- h). In this work we study the dependence of its fractal dimension on the system parameters. We also examine the time required for the system to fill the entire space in the limit $M \rightarrow \infty$. Finally, we observe oscillations in the wavefront, which arise from the delay between the active–immune and immune–sensitive transitions.

\section{Fractal dimension of the wavefront}
The wavefront fractal dimension is defined as 
\begin{equation} \label{fr dim}
    D = \lim_{L\rightarrow\infty}\dfrac{\log{B}}{\log{L}},
\end{equation}

where $B$ is the number of active and immune cells which have a sensitive neighbor (wavefront border), and $L$ is the automaton size in cell units.\\

\begin{figure}[H] 
    \centering
    \subfloat[]{\includegraphics[scale = 0.25]{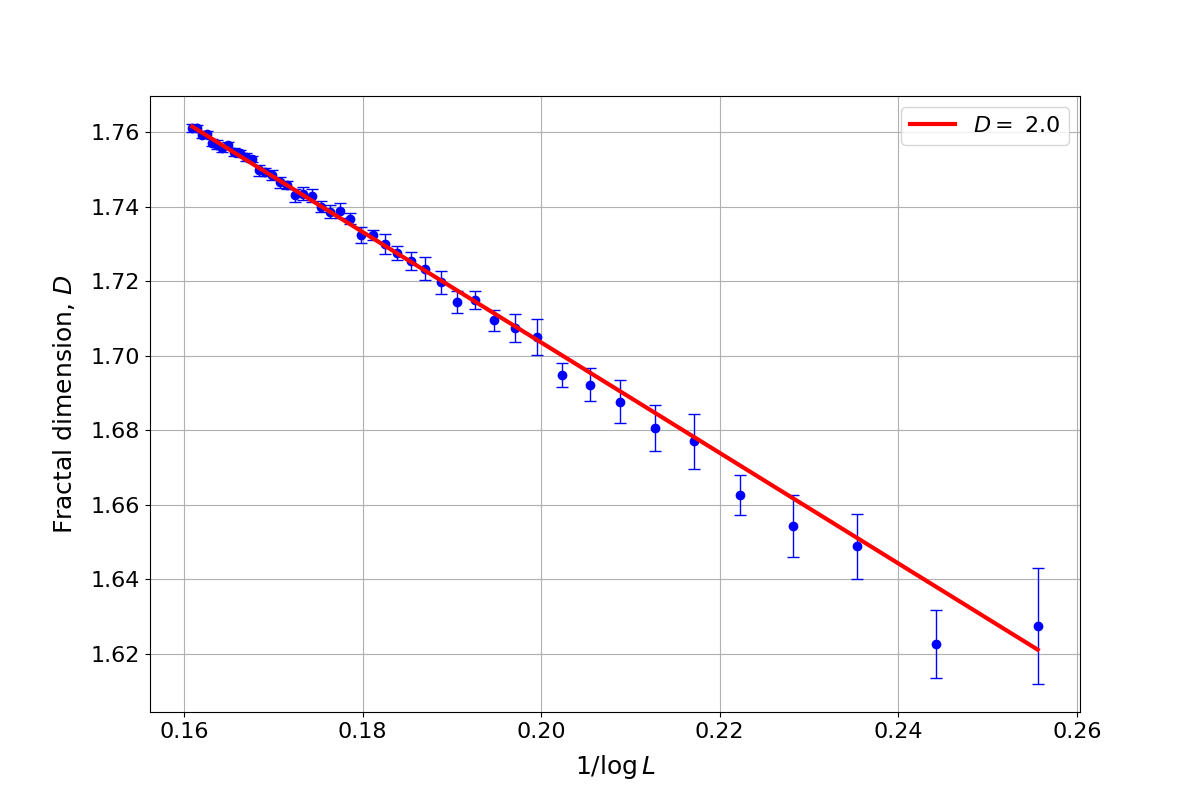}}
    \centering
    \subfloat[]{\includegraphics[scale = 0.25]{Dim_on_1logNp=0.7_n=10.png}}\\
    \centering
    \subfloat[]{\includegraphics[scale = 0.25]{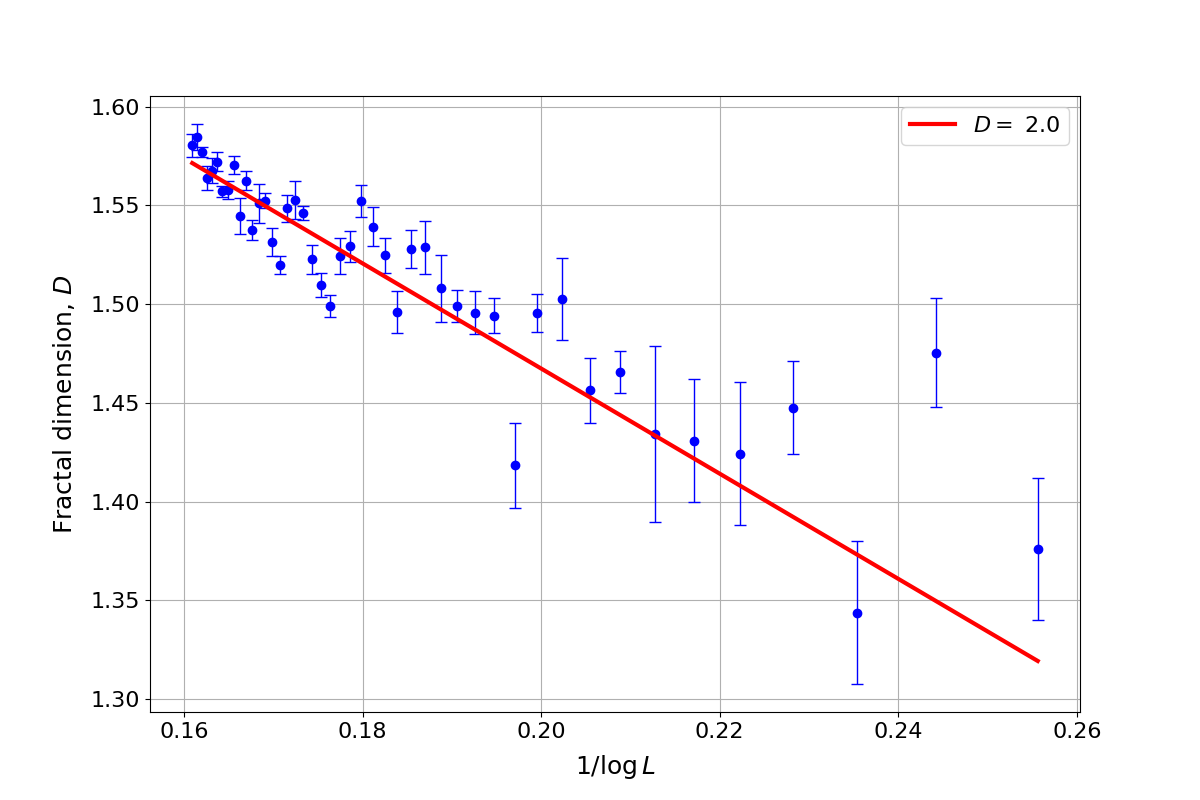}}
    \centering
    \subfloat[]{\includegraphics[scale = 0.25]{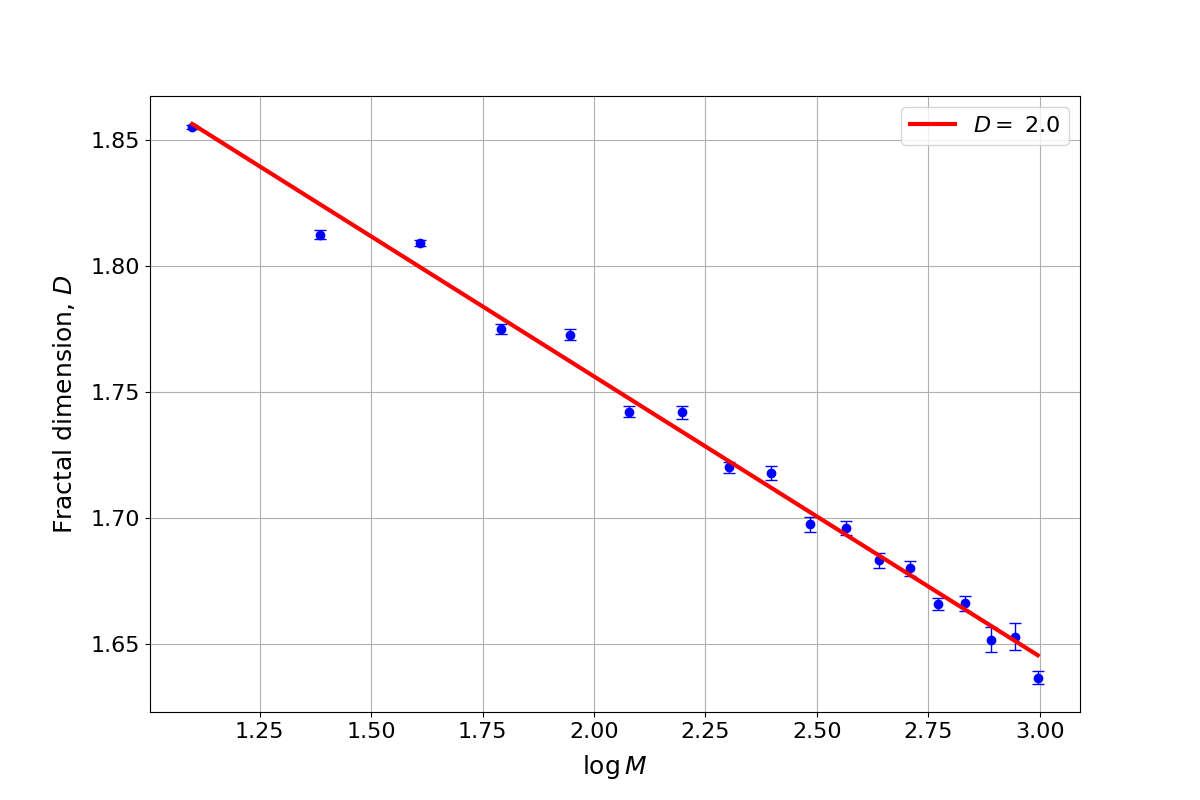}}
    \caption{The fractal dimension limit, \ref{D of L}a –- the dimension dependence on automaton size $G = 0.7$, $M = 10$, \ref{D of L}b -– dependence on $1/\log{L}$ $G = 0.7$, $M = 10$, \ref{D of L}c --  $G = 0.44$, $M = 5$, \ref{D of L}d -- the dimension dependence on $\log{M}$ with fixed $L = 200$, $G=0.7$.}
    \label{D of L}
\end{figure}

We first verify the existence of the limit. Figures \ref{D of L}a--c show that the dimension approaches a limiting value as $1/log{L}$. The fractal dimension depends neither on the activation probability for $G > 0.44$ nor on the immune tail length $M$. Note, however, that for a fixed system size $L$, the wavefront length decreases with increasing immune tail length. This behavior arises because the immune size defines the characteristic wavelength of the activity waves, establishing a scale that is compared against the system size. Thus, increasing the immune tail length $M$ is equivalent to reducing the automaton size $L$. As a result, the limit (\ref{fr dim}) corresponds to the limit $M\rightarrow 0$ (Fig. \ref{D of L}d).\\

The fractal dimension equals 2 for $G>0.44$. The activity is distributed in such a way as to maximize the number of excitable cells. Within a finite volume, the system self-organizes into a dynamic equilibrium. The immune walls generated by active cells prevent activity from reaching its maximum. Consequently, the ratio $\log{B}/\log{L}$ remains strictly less than 2 for any finite $L$ and $M$ (Fig. \ref{D of L}).\\

\begin{figure}[H] 
    \centering
    \includegraphics[scale = 0.5]{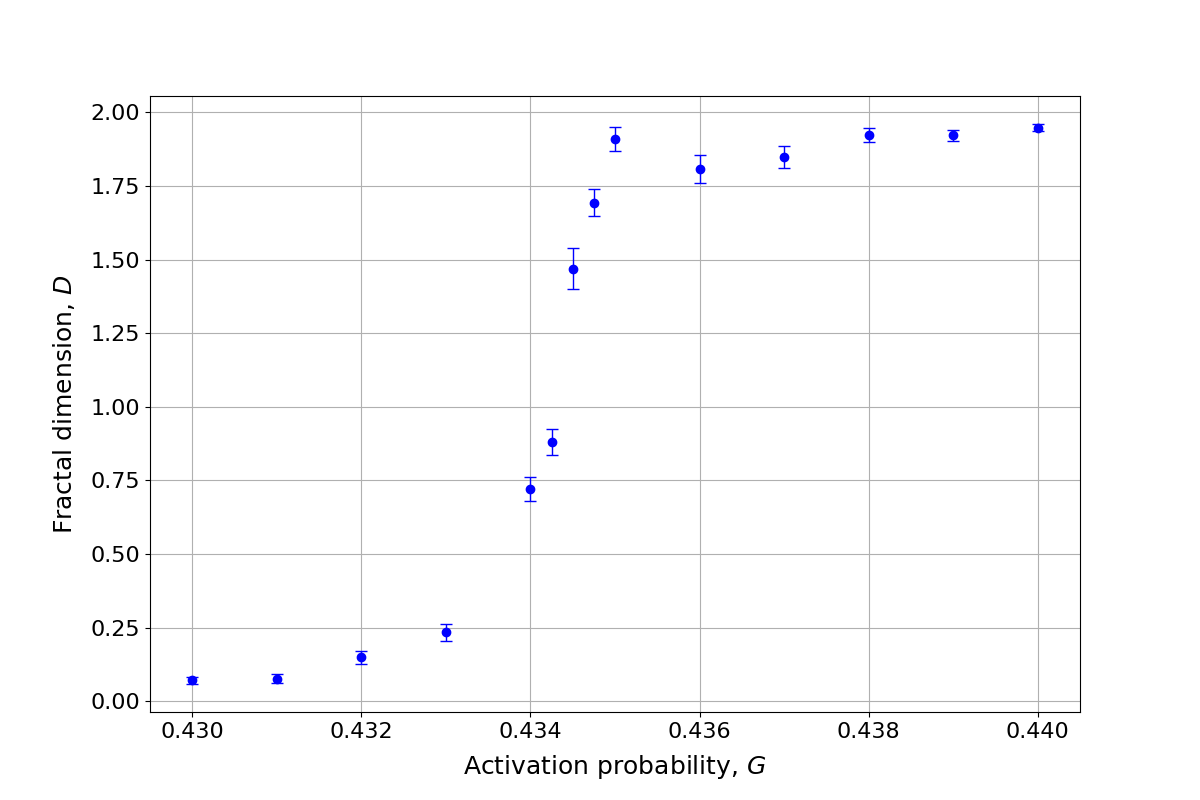}
    \caption{The dependence of the fractal dimension on the activation probability near critical point.}
    \label{Dim of G}
\end{figure}
The fractal dimension exhibits a rapid transition near the critical point, taking a value of zero for activation probabilities below the critical threshold and two in the overcritical regime. This shift is confined to a narrow interval, $0.43 <G<0.44$ (Fig. \ref{Dim of G}), where the system switches from the fading mode to the wave mode.

\section{Fluctuation statistic}
\begin{figure}[H] 
    \centering
    \includegraphics[scale = 0.5]{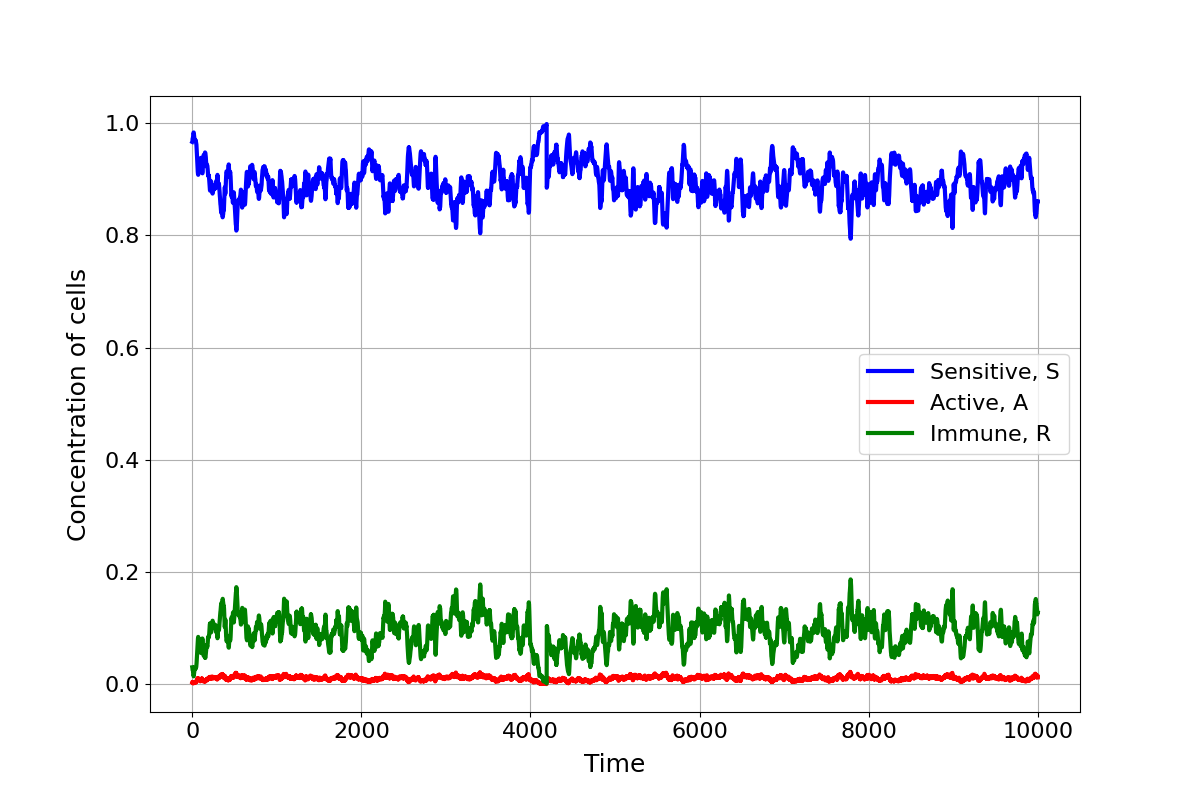}
    \caption{The cells concentration evolution $G = 0.47$, $M = 10$.}
    \label{SIR of T}
\end{figure}

\begin{figure}[H] 
    \centering
    \subfloat[]{\includegraphics[scale = 0.25]{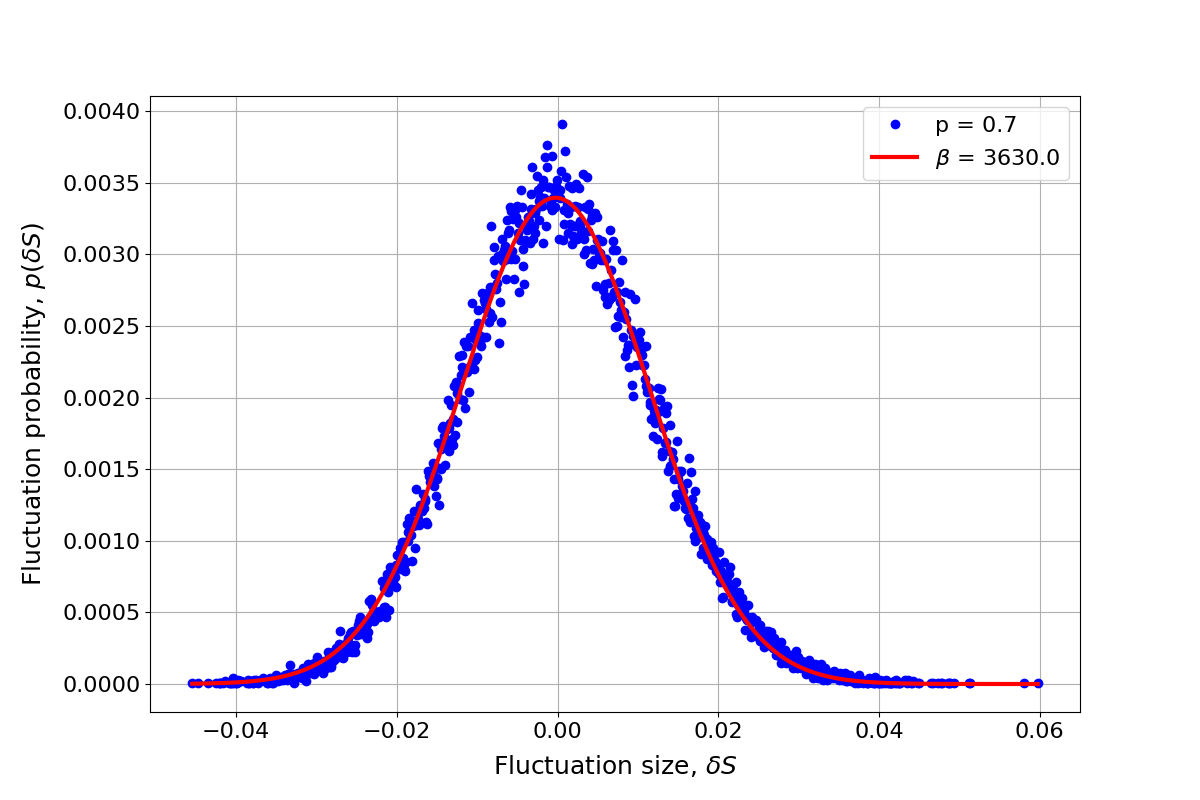}}
    \centering
    \subfloat[]{\includegraphics[scale = 0.25]{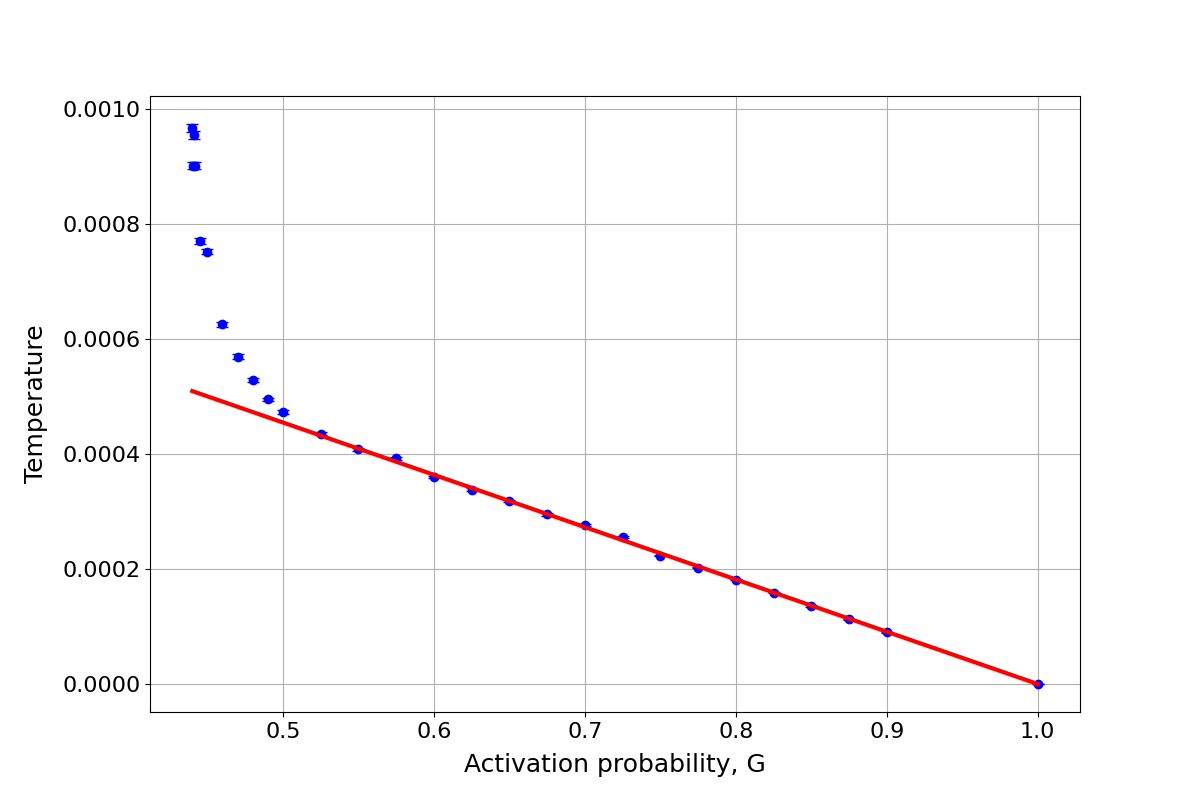}}
    \\
    \centering
    \subfloat[]{\includegraphics[scale = 0.25]{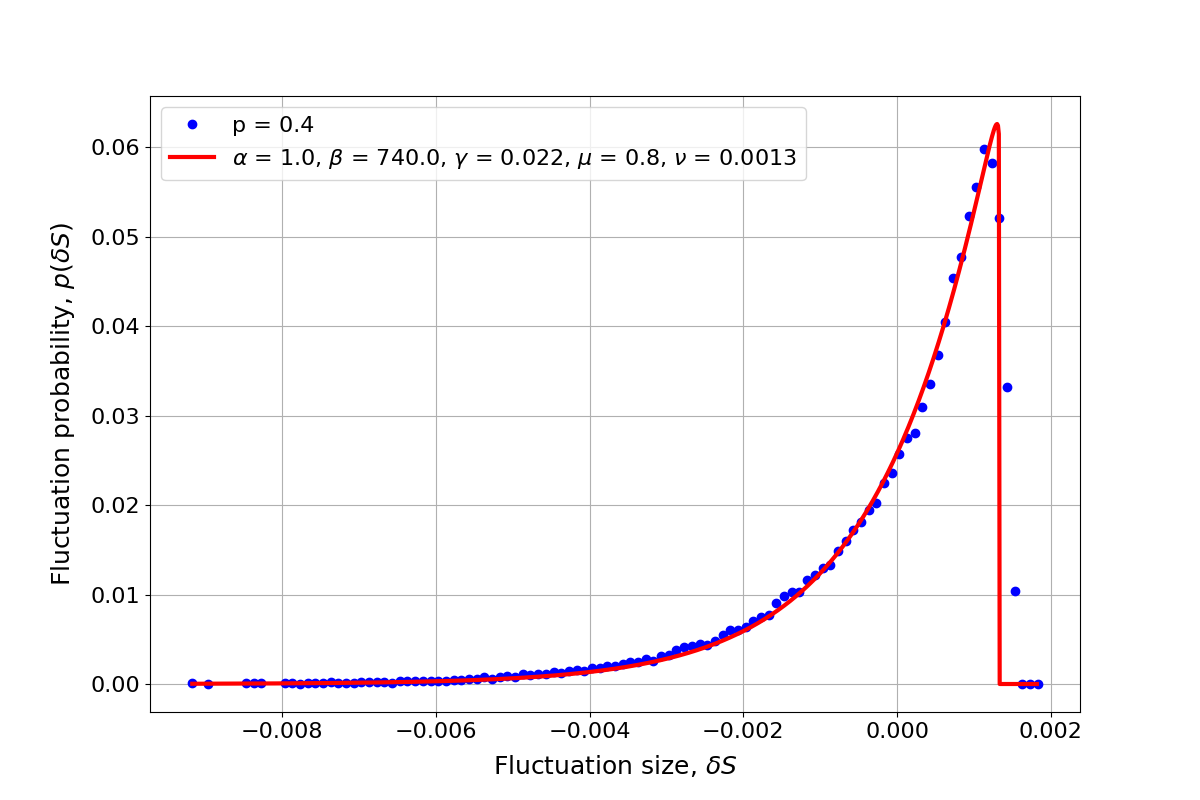}}
    \centering
    \subfloat[]{\includegraphics[scale = 0.25]{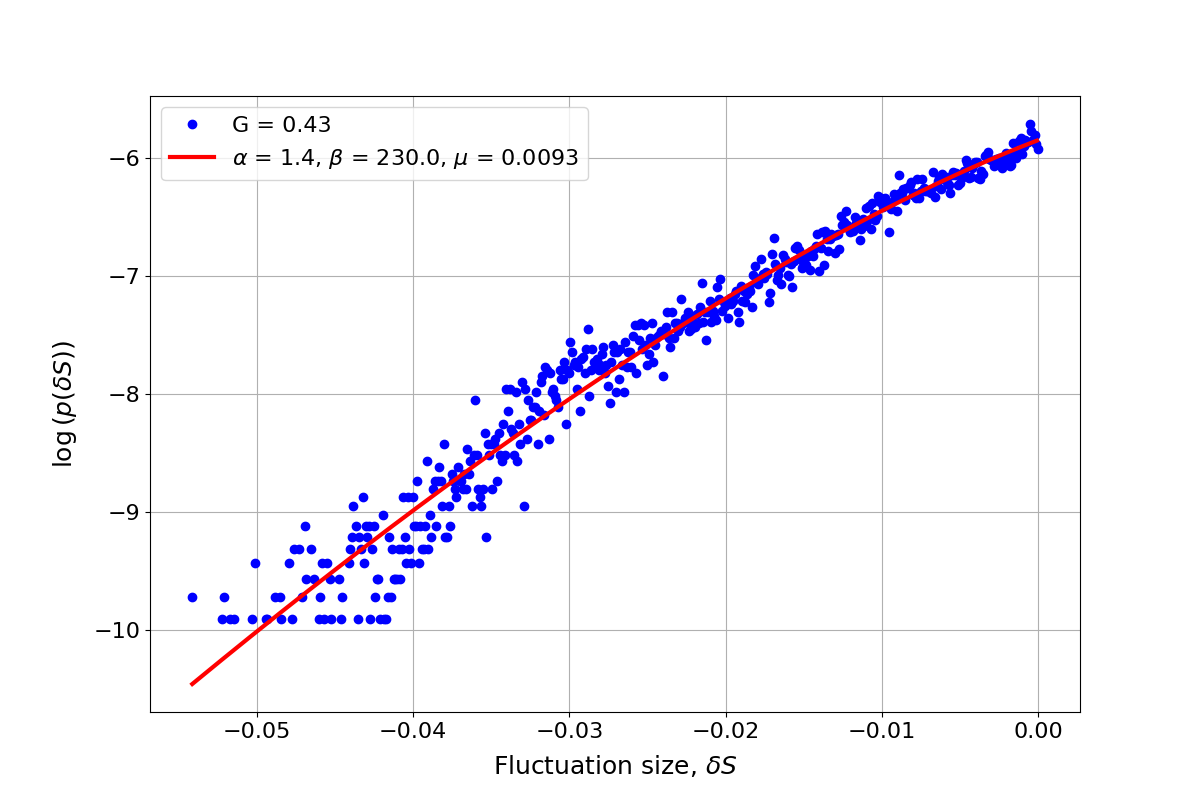}}
    \caption{The fluctuation distribution $M = 5$: \ref{temperature}a -- the Gauss approximation of the distribution in the wave mode $G = 0.7$, \ref{temperature}b -- the dependence system temperature on $G$, \ref{temperature}c -- distribution in the fading mode $G = 0.4$, \ref{temperature}d -- the Weibull approximation of distribution near critical point $G = 0.43$}
    \label{temperature}
\end{figure}

Similar behavior are also evident in the probability distribution of fluctuations in the number of sensitive cells. All of them follow a Gaussian distribution for $G > 0.44$ (Fig. \ref{SIR of T}, \ref{temperature}a) 
\begin{equation}
    p(\delta S) = \dfrac{1}{Z}\exp{(-\beta (\delta S)^2)},
\end{equation}
where $S$ is the concentration of sensitive cells, $\delta S = S - \bar{S}$ is the fluctuation in the concentration of sensitive cells, $\bar{S}$ is the mean concentration of sensitive cells. We assume that the variance of the fluctuations represents the temperature of the system ($T = 1/\beta$). The temperature then decreases with increasing activation probability and follows a linear law on the interval $0.5 < G < 1$ (Fig. \ref{temperature}b).

\begin{figure}[H] 
    \centering
    \subfloat[]{\includegraphics[scale = 0.25]{Fluctuation_statistic_minus_tail_N=100_p=0.43_n=5_VolS=100000.png }}
    \centering
    \subfloat[]{\includegraphics[scale = 0.25]{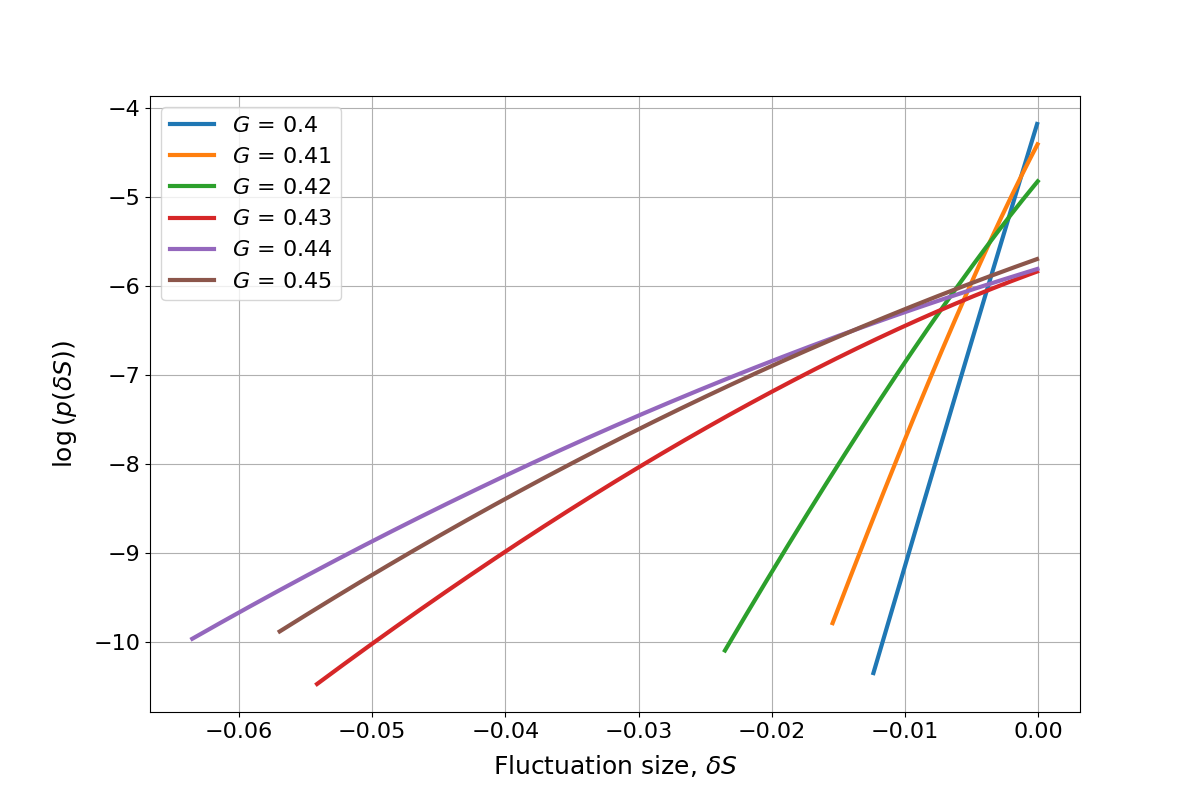}}\\
    \centering
    \subfloat[]{\includegraphics[scale = 0.25]{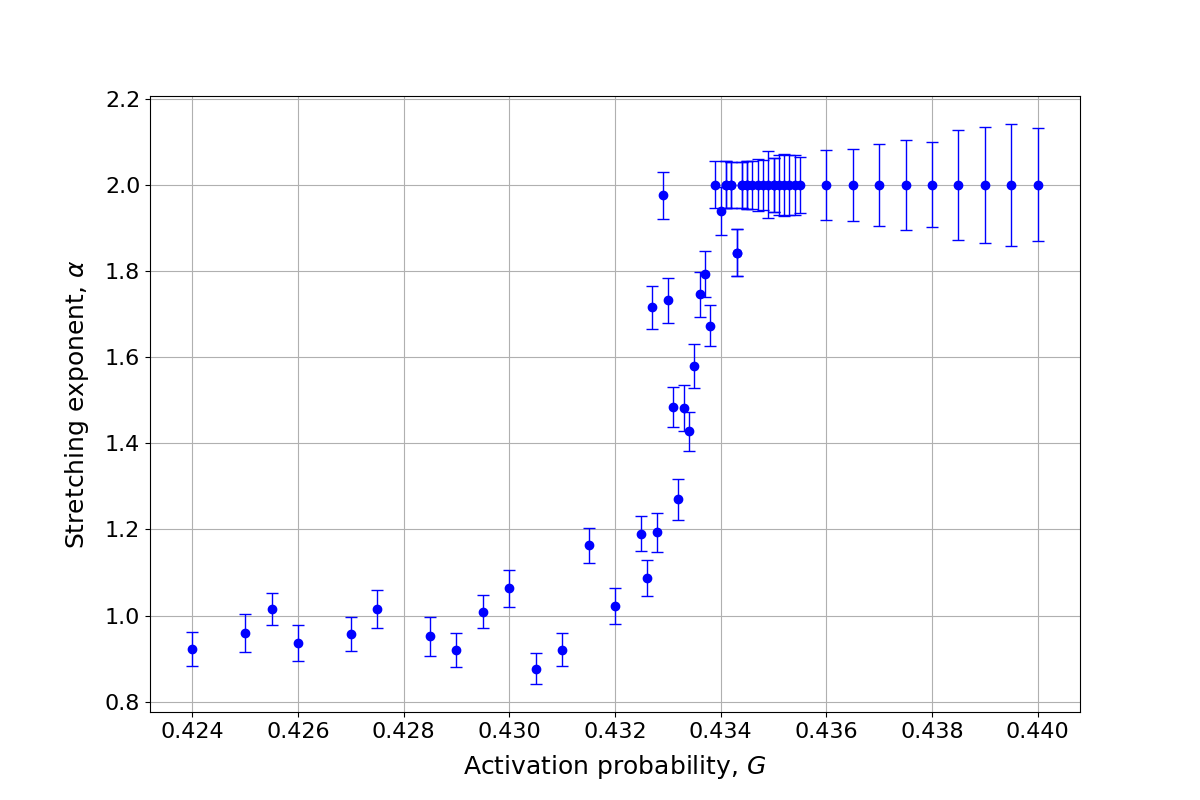}}
    \centering
    \subfloat[]{\includegraphics[scale = 0.25]{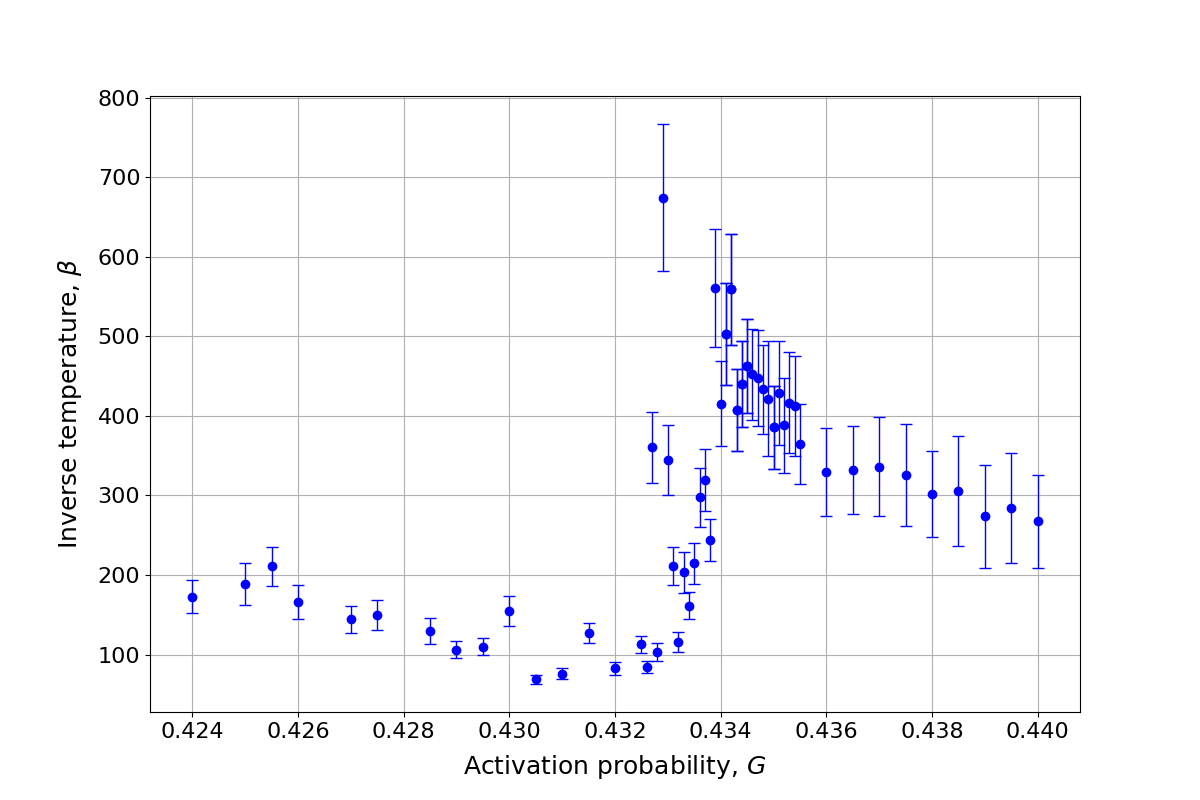}}
    \caption{The distribution of negative fluctuations $M = 5$: \ref{Minus tail}a -- $G = 0.43$, \ref{Minus tail}b -- the distribution transformation from exponential to Gaussian, \ref{Minus tail}c -- dependence of the stretching exponent on the activation probability, \ref{Minus tail}d -- dependence of the inverse temperature on the activation probability.}
    \label{Minus tail}
\end{figure}

For $G<0.44$, the distribution departs from normality and acquires a pronounced asymmetry (Fig. \ref{temperature}c, d). The negative tail exhibits a stretched-exponential decay, $\sim\exp{(-(\delta S)^\alpha)}$, with the stretching exponent $\alpha$ ranging from 1 (recovered at low $G$) up to 2 as the activation probability grows (Fig. \ref{Minus tail}a, b). Conversely, positive fluctuations are severely limited in magnitude. Their distribution is non-exponential, and the probability density function attains its maximum within the domain of small positive fluctuations.
We approximate the probability density function (Fig. \ref{temperature}a, c, d) as
\begin{gather} \label{gen dist}
    p(\delta S) = 
    \begin{cases}
        \dfrac{1}{Z}(\nu-\delta S)^\gamma\exp{(-\beta|\delta S-\mu|^\alpha)} \qquad \delta S \leq \nu\\
        0 \qquad \delta S>\nu
    \end{cases},
\end{gather}
where $\alpha$ is a fitting parameter corresponding to the stretching exponent, $\beta$ is the inverse temperature of the system, $\gamma$ controls how rapidly the density vanishes as $\delta S$ approaches the upper boundary $\nu$, $\mu$ is parameter of exponent asymmetry, $\nu$ is the upper bound on positive fluctuations, and $Z$ is the normalization constant. As a zeroth-order approximation, the parameters $\alpha$, $\beta$, and $\mu$ are extracted from a fit to the tail of negative fluctuations (Fig. \ref{Minus tail}a, b), which itself is approximated by
\begin{gather}
    p(\delta S) \approx \dfrac{1}{Z}\exp{(-\beta|\delta S-\mu|^\alpha)}.
\end{gather}

This approximation illustrates how the statistics of negative fluctuations evolve across a phase transition. In the fading regime $(G<G_{cr})$, the fluctuation distribution is exponential, whereas in the wave regime $(G>G_{cr})$, it becomes Gaussian. Near the critical point $(G\approx G_{cr})$, a sharp transition takes place, linking the exponential and Gaussian statistics (Fig. \ref{Minus tail}).\\

Different parameters become irrelevant in different regimes. In the overcritical regime, $\mu = 0$ and $\nu \rightarrow \infty$; here either $\gamma$ has almost no effect on the fit, or one may set $\gamma = 0$ in which case the fit becomes independent of $\nu$. In the fading regime $\alpha \rightarrow 1$ and $\mu$ enters only through the normalization constant, provided $\mu > \nu$. In this regime $\gamma$ is small, which ensures a sharp cutoff of positive fluctuations. All parameters become essential near the critical point where the two regimes meet. A particularly interesting case is $G = 0.43$. There the density function reduces to a Weibull distribution \cite{Weibull} (Fig. \ref{temperature}d)
\begin{gather}
    p(\delta S) = \dfrac{1}{Z}\left(\nu -\delta S\right)^{\alpha -1}\exp{\left(-\beta\left(\nu-\delta S\right)^\alpha\right)}, \qquad \gamma \approx \alpha -1, \qquad \mu \approx \nu.
\end{gather}

\begin{figure}[H] 
    \centering
    \includegraphics[scale = 0.5]{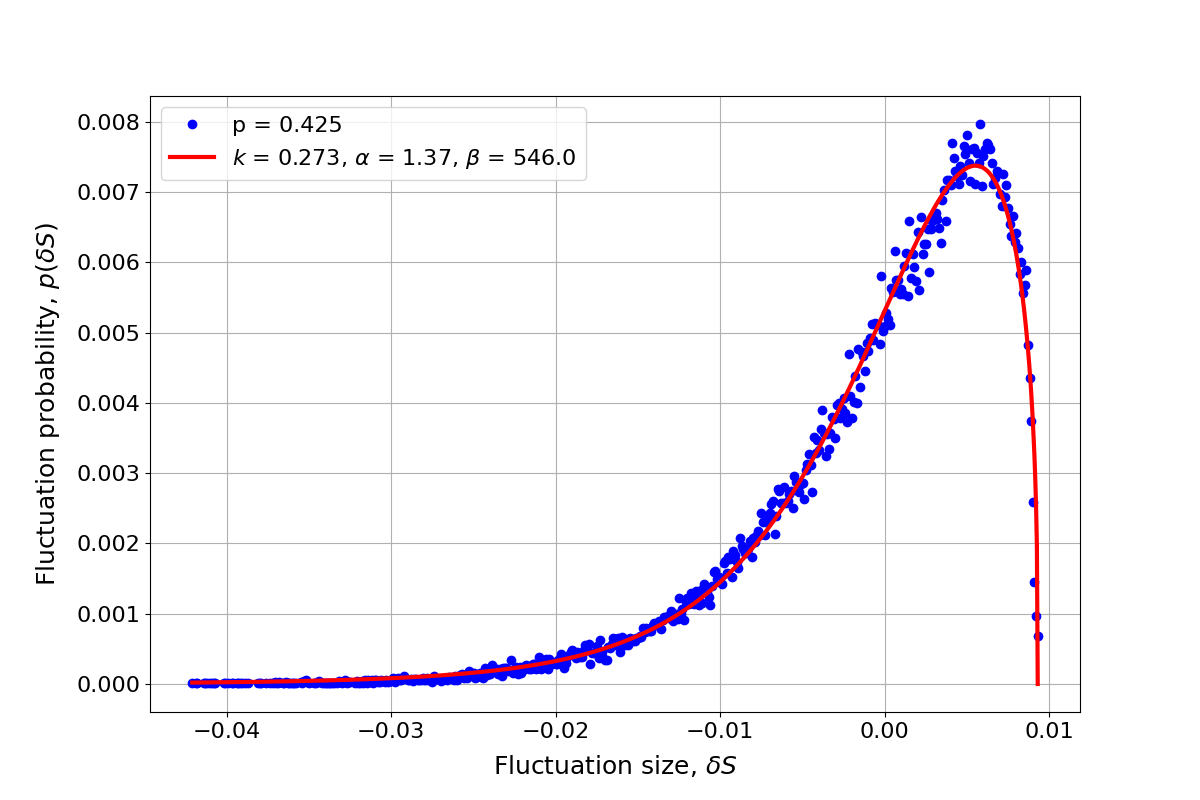}
    \caption{The $\kappa$-Weibull approximation of the critical fluctuations distribution $G= 0.425, M = 5$.}
    \label{k-Weibull}
\end{figure}

This suggests that near the critical point in the fading mode, the fluctuation probability density function is a generalization of the Weibull distribution. A related distribution, the Kaniadakis-Weibull distribution, has been observed in epidemiological studies \cite{Gen Stat}, \cite{Kaniadakis}. Its probability density function is given by
\begin{gather}
    p(x) = 
    \begin{cases}
        \dfrac{\alpha \beta x^{\alpha -1}}{\sqrt{1 + \kappa^2\beta^2x^{2\alpha}}} \exp_\kappa{\left(-\beta x^\alpha\right)} \qquad x\geq0\\
        0 \qquad x < 0
    \end{cases},
\end{gather}
where $\exp_\kappa{x} = (\sqrt{1+\kappa^2x^2}+\kappa x)^{1/\kappa}$, $x = \nu - \delta S$. Fig.\ref{k-Weibull} demonstrates how the probability density function is approximated by this $\kappa$-Weibull density. This implies that negative fluctuations exhibit the asymptotic power-law tail $\sim |\delta S|^{-(1+\alpha/\kappa)}$.\\

\begin{figure}[H] 
    \centering
    \subfloat[]{\includegraphics[scale = 0.25]{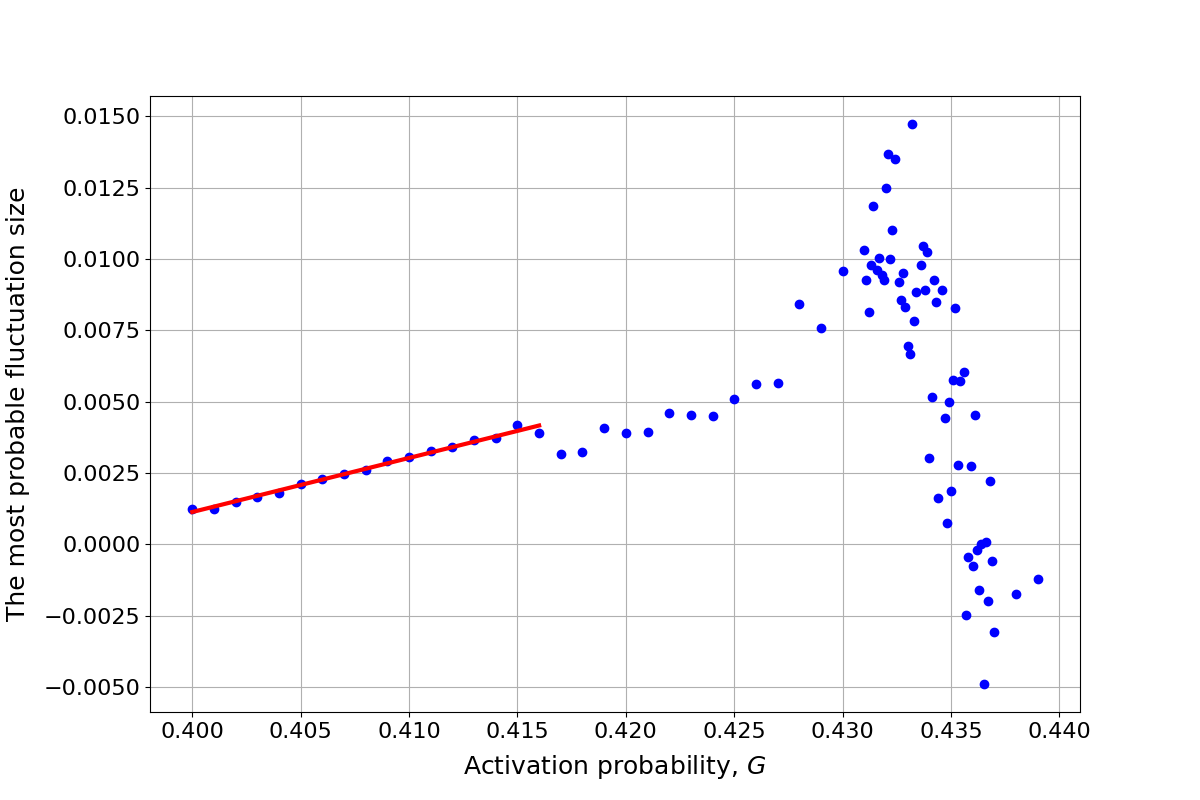}}
    \centering
    \subfloat[]{\includegraphics[scale = 0.25]{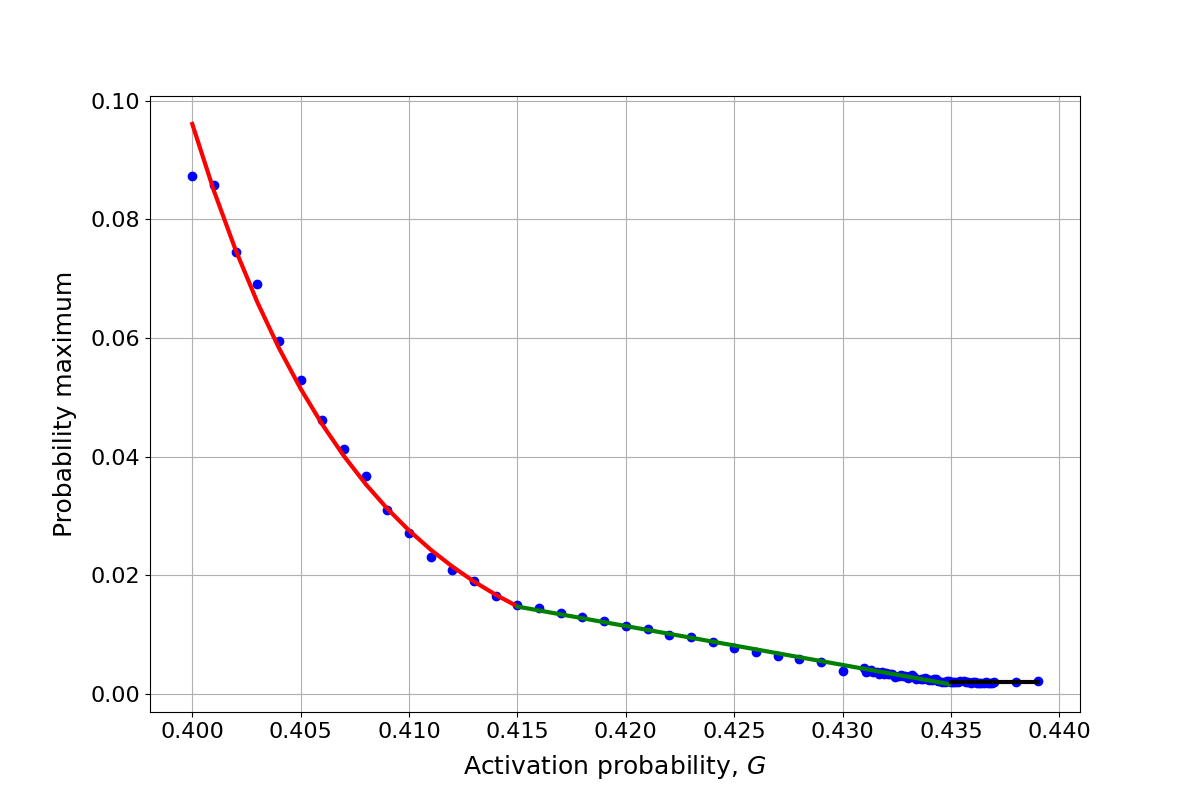}}
    \\
    \centering
    \subfloat[]{\includegraphics[scale = 0.25]{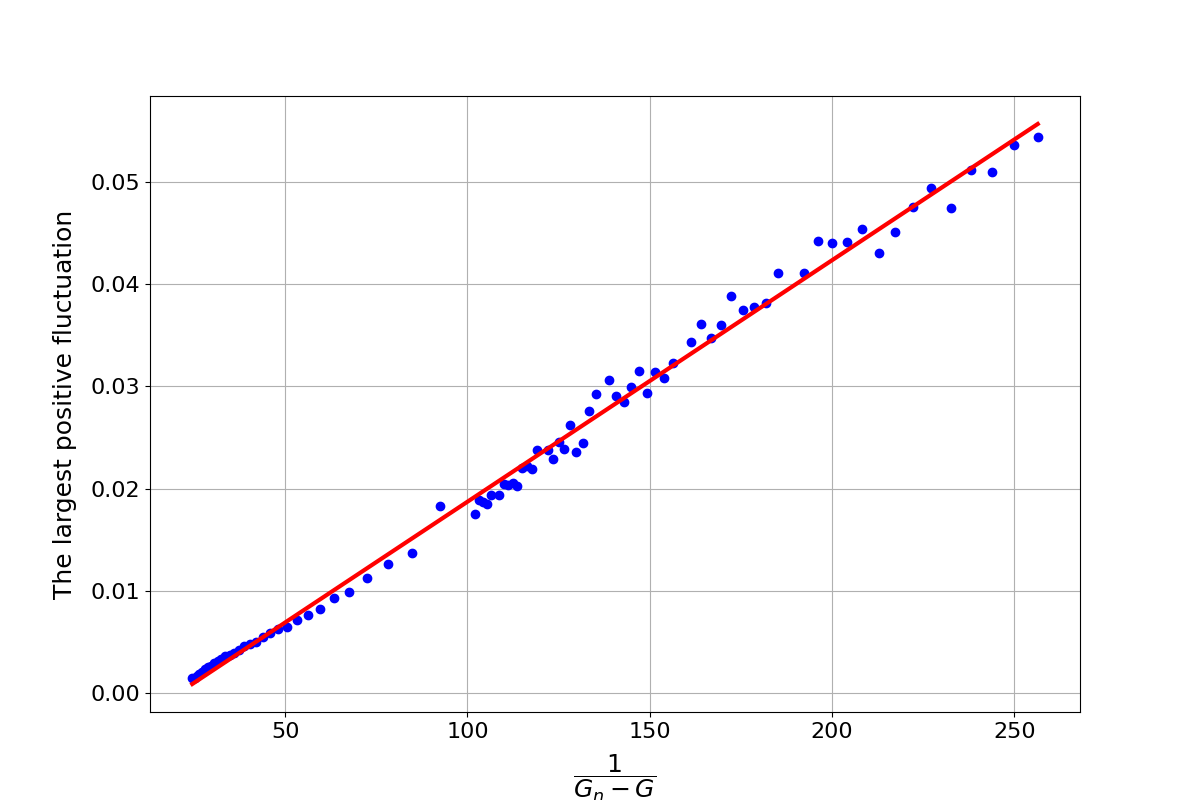}}
    \centering
    \subfloat[]{\includegraphics[scale = 0.25]{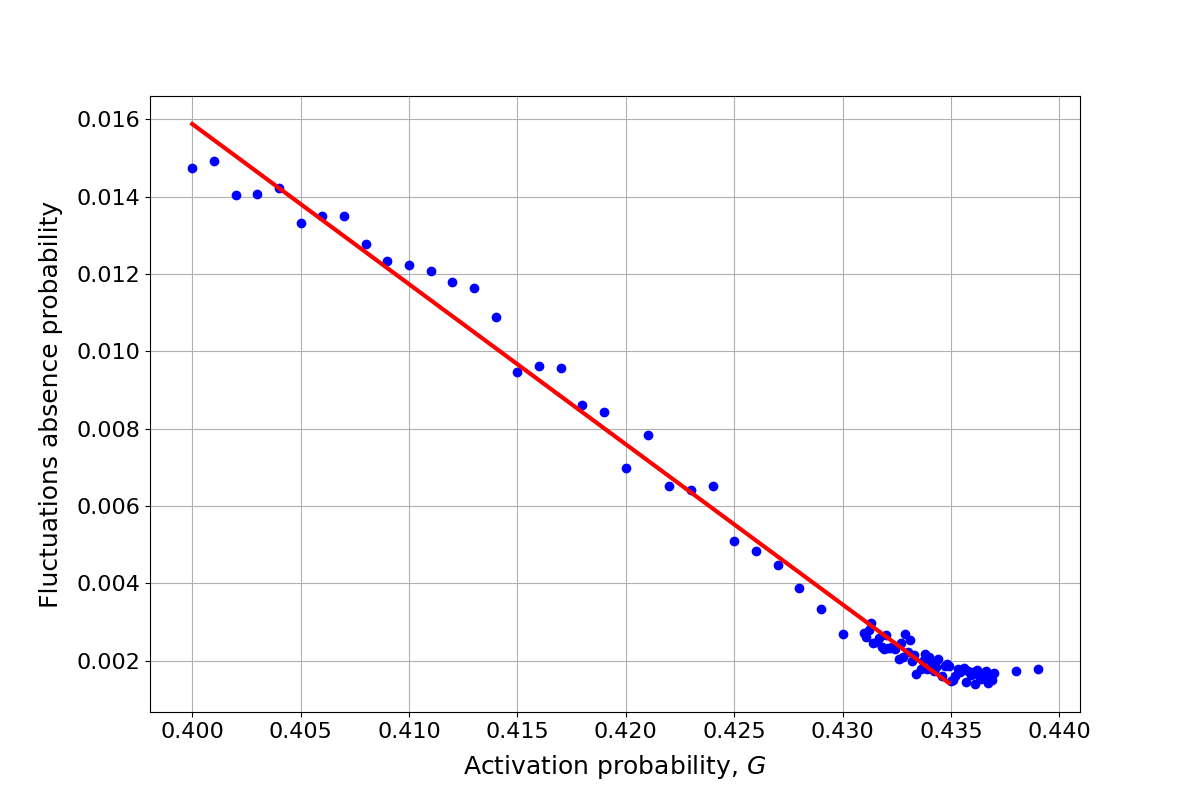}}
    \caption{Properties of the fluctuation distribution for $M = 5$: \ref{property} -- the most probable fluctuation size; \ref{property}b -- the maximum of the fluctuation distribution (red line — exponential approximation, green line — linear, black line — constant); \ref{property}c -- the largest positive fluctuation, where $G_n \approx 0.4408$ is the activation probability above which the distribution becomes normal; \ref{property}d -- the probability of no fluctuations.}
    \label{property}
\end{figure}

We examine how the probability density function of fluctuations depends on the activation probability, focusing on several key characteristics: the most probable fluctuation size (Fig. \ref{property}a), its corresponding probability (Fig. \ref{property}b), the magnitude of the largest possible positive fluctuation (Fig. \ref{property}c), and the probability of observing no fluctuation (Fig. \ref{property}d). Over certain parameter ranges, these dependencies is  transformed to a linear form near the critical point.\\

In the fading regime ($G<0.417$), the most probable fluctuation size increases linearly with activation probability. At the critical point, this quantity exhibits a sharp transition and then drops to zero, consistent with a Gaussian distribution. The probability maximum decays as $\sim\exp(−aG)$ within the fading regime (Fig. \ref{property}b); closer to the critical point, the decay becomes linear, and at the critical point itself, it saturates to a constant value. Beyond the critical point, its behavior follows the temperature dependence of the Gaussian distribution (Fig. \ref{temperature}b).\\
The largest positive fluctuations grow as $\sim1/(G_n−G)$ near the critical point, where $G_n \approx 0.4408$ denotes the activation probability beyond which the fluctuation distribution becomes strictly Gaussian and unbounded from above (Fig. \ref{property}c). Meanwhile, the probability of $\delta S=0$ decreases linearly with activation probability (Fig. \ref{property}d). At $G\approx G_n$, it reaches a constant value equal to the probability maximum; beyond this point, both quantities simply coincide.\\

These simple linear dependencies can be used to simplify approximations of the fluctuation probability density functions or to check their consistency. If we approximate the distribution using the six-parameter function (\ref{gen dist}) and take the specified dependencies into account, we are left with only two independent parameters. The largest positive fluctuation determines $\nu$. The remaining parameters are governed by the following relations:
\begin{gather}
    \gamma = \alpha \beta(\mu-(\delta S)_{m})^{\alpha - 1}(\nu - (\delta S)_{m}), \label{gamma}
\end{gather}
\begin{gather}
    Z = \dfrac{\nu^\gamma}{p_0}\exp{(-\beta \mu^\alpha)} \label{Z},
\end{gather}
\begin{gather}
    \mu^\alpha - (\mu - (\delta S)_{m})^\alpha+\alpha (\nu - (\delta S)_{m})\log{\left(1-\dfrac{(\delta S)_m}{\nu}\right)}(\mu - (\delta S)_{m})^{\alpha-1} = \dfrac{1}{\beta}\log{\dfrac{p_m}{p_0}} \label{mu},
\end{gather}
where $(\delta S)_m$ is the most probable fluctuation size, $p_m$ is its probability, $p_0$ is the probability of no fluctuations. Given $\alpha$ and $\beta$, we can solve (\ref{mu}) numerically for $\mu$. Then, $\gamma$ follows from (\ref{gamma}), and $Z$ is obtained from (\ref{Z}). Consequently, the fitting procedure requires only two free parameters, which can be estimated by approximating the tails of the negative fluctuations (Fig. \ref{Minus tail}c--d).\\

\section{Conclusions}
In this article, we investigate the phase transition between exponential and Gaussian-distributed fluctuations. Near the critical point, fluctuations have a $\kappa$-Weibull distribution, which serves as a bridge between the two phases.\\

This transition also corresponds to a change in the fractal dimension of the wavefront from zero to two. This is in line with a qualitative understanding of the system's behavior. For $G<G_{cr}$, the system fades and the wavefront shrinks to points whose fractal dimension is zero. On the other hand, for $G > G_{cr}$, the automaton evolution consists of all-filling stochastic waves whose fractal dimension is two. Only in a narrow interval around $G_{cr}$ does the wavefront have a fractal dimension between zero and two.\\

We thus establish a relationship between fractal dimension and fluctuation statistics for this system. Isolated activity centers, characterized by a fractal dimension of zero, exhibit an exponential distribution. Conversely, all-filling waves, with a dimension of two, follow a Gaussian distribution. In the intermediate, critical regime, complex avalanches display dimensions between zero and two, and their distribution is described by the $\kappa$-Weibull.    
\section{Acknowledgments}
The work of M.Yu.Satleikin was supported by the Theoretical Physics and Mathematics Advancement Foundation “BASIS” (grant № 25-2-1-74-1).

\end{document}